\documentclass{PoS}

\title{Charm Quark Mass with Calibrated Uncertainty}

\ShortTitle{Charm Quark Mass}

\author{\speaker{Jens Erler}\\
        Departamento de F\'isica Te\'orica \\
        Instituto de F\'isica \\
        Universidad Nacional Aut\'onoma de M\'exico \\
	Apartado Postal 20--364, M\'exico CDMX 01000, M\'exico \\
        E-mail: \email{erler@fisica.unam.mx}}

\author{Pere Masjuan\\
        Grup de F\'{\i}sica Te\`orica, Departament de F\'{\i}sica, Universitat Aut\`onoma de Barcelona \\
	{\rm and} \\
	Institut de F\'{\i}sica d'Altes Energies (IFAE) \\
	The Barcelona Institute of Science and Technology (BIST) \\
	Campus UAB, E--08193 Bellaterra (Barcelona), Spain \\
        E-mail: \email{masjuan@ifae.es}}

\author{Hubert Spiesberger\\
        PRISMA Cluster of Excellence, Institut f\"ur Physik \\
       Johannes Gutenberg-Universit\"at, 55099 Mainz, Germany \\
        E-mail: \email{spiesber@uni-mainz.de}}

\abstract{We determine the charm quark mass $\hat m_c(\hat m_c)$ 
from QCD sum rules of moments of the vector current correlator calculated in perturbative QCD. 
Only experimental data for the charm resonances below the continuum threshold are needed in our approach, 
while the continuum contribution is determined by requiring self-consistency between various sum rules, including the one for the zeroth moment. 
Existing data from the continuum region can then be used to bound the theoretical error. 
Our result is $\hat m_c(\hat m_c) = 1272 \pm 8$~MeV for $\hat\alpha_s(M_Z) = 0.1182$. 
Special attention is given to the question how to quantify and justify the uncertainty.}

\FullConference{EPS-HEP 2017, European Physical Society conference on High Energy Physics\\
		5-12 July 2017\\
		Venice, Italy \\ 
		\vspace{24pt} 
		{\rm MITP/17--046}}

\begin{document}

\section{Method and result}

We report on work~\cite{Erler:2016atg} on the determination of the charm quark mass, $m_c$.
First, let us recall why precise values of $m_c$ are needed.
Besides being a fundamental input parameter defining the Standard Model (SM), 
it enters many QCD processes, as we all as a number of electroweak quantities.  
For example, it is needed for the renormalisation group running of the fine structure constant 
from the Thomson limit to the $Z$ pole~\cite{Erler:1998sy}.
Incidentally, the physics involved in this running is the basis of what we will refer to as the $0^{\rm th}$ moment of the sum rule later on. 
Conversely, if one wants to run the weak mixing angle from the $Z$ pole down to low energies one can do so in perturbation theory as well,
but one needs an input value for $m_c$~\cite{Erler:2004in}. 
Finally, $m_c$ enters the SM prediction of the anomalous magnetic moment of the muon 
when the heavy quark contribution is evaluated perturbatively~\cite{Erler:2002bu}.
The most important application in the future will be the test of the mass {\em versus\/}\ Yukawa coupling relation in the single Higgs SM,
because at future lepton colliders it will be possible to measure the Yukawa coupling of the charm quark to very high precision.
If one --- within the SM -- converts the currently projected precision to a mass measurement this would correspond to an uncertainty of 
about 8~MeV, which should then be the benchmark of what one wants to achieve at least regarding the precision in $m_c$. 
Of course, one can determine $m_c$ in lattice simulations, 
but it is prudent to inquire for a second opinion derived from an independent first principles approach.

This approach is provided by the relativistic QCD sum rule formalism describing low moments of the charm vector-current correlator $\Pi_c$,
where the master equation is given by
\begin{equation}
\label{SR}
12 \pi^2\, \frac{\Pi_c (0) - \Pi_c (-t)}{t} = \int_{4 \hat m_c^2}^\infty \frac{{\rm d} s}{s} \frac{R_c(s)}{s + t}\ .
\end{equation}
The right-hand side is basically an integral over electromagnetic charm pair production 
(normalized to the muonic cross-section) with certain weights.
$\Pi_c$ has been calculated in perturbative QCD to 
${\cal O}(\alpha_s^3)$~\cite{Chetyrkin:2006xg,Boughezal:2006px,Kniehl:2006bf,Maier:2008he,Maier:2009fz}.
In the limit $t \to 0$, the left-hand side of Eq.~(\ref{SR}) turns into a derivative
with the right-hand side being suppressed by two powers of $s$ and providing the $1^{\rm st}$ moment, ${\cal M}_1$.
Taking further derivatives generates higher moments, ${\cal M}_n$~\cite{Novikov:1977dq}.
But one can also take the opposite limit, $t \to \infty$, which corresponds to the $0^{\rm th}$ moment sum rule, 
${\cal M}_0$~\cite{Erler:2002bu}, mentioned already before.
As it stands, ${\cal M}_0$ would be divergent and needs regularization, but this can be done {\em e.g.}, 
by subtracting the asymptotic perturbative expansion, $R_c(s) \equiv 4/3 \lambda_1(s)$ at $m_c = 0$, on both sides.

${\cal M}_0$ is one of the ingredients where our analysis differs from others. 
Another special feature of our approach is that the only experimental input are the electronic widths of the $J/\psi$ and the $\psi(2S)$
narrow vector resonances. 
The continuum contribution will be constrained by requiring self-consistency between various sum rules. 
This is where ${\cal M}_0$ comes in handy because it experiences stronger sensitivity to the continuum compared to higher moments,
and at the same time a milder sensitivity to $m_c$, so that it provides a useful handle. 
This means that quark-hadron duality is needed only in the {\em finite\/} region between the $\psi(2S)$ and the onset of open charm production.
By contrast, in previous analyses one performs the integral in Eq.~(\ref{SR}) up to some value of $s$ after which no more data are available 
and where one needs to switch to perturbative QCD.
This involves changing from hadron to quark degrees of freedom at this particular point which is not rigorously justified.
Finally, in our approach it is also possible to estimate correlated errors across various moments.
Our result for the $\overline{\rm MS}$ charm quark mass is
\begin{equation}
\hat m_c(\hat m_c) = 1272 \pm 8 + 2616 [\hat\alpha_s(M_Z) - 0.1182]~{\rm MeV},
\label{cmresult}
\end{equation} 
where the theoretical and experimental uncertainties add up to 8~MeV, coinciding with the benchmark target mentioned earlier.
In addition, we explicitly display the parametric dependence on the strong coupling constant, $\hat\alpha_s(M_Z)$.
For this particular result --- the main result of our work --- we used ${\cal M}_0$ in tandem with ${\cal M}_2$, 
where we assumed the error component associated with the truncation of perturbative QCD to be uncorrelated between the two
(we expect this error contribution to be stronger correlated among the higher moments).
The central value is in very good agreement with other recent sum rule determinations but there is less agreement regarding 
the theory dominated uncertainty, so we wish to go into a few more details about it.

\section{Evaluation of the uncertainty}

\begin{figure}[b]
\centering
\includegraphics[width=.6\textwidth]{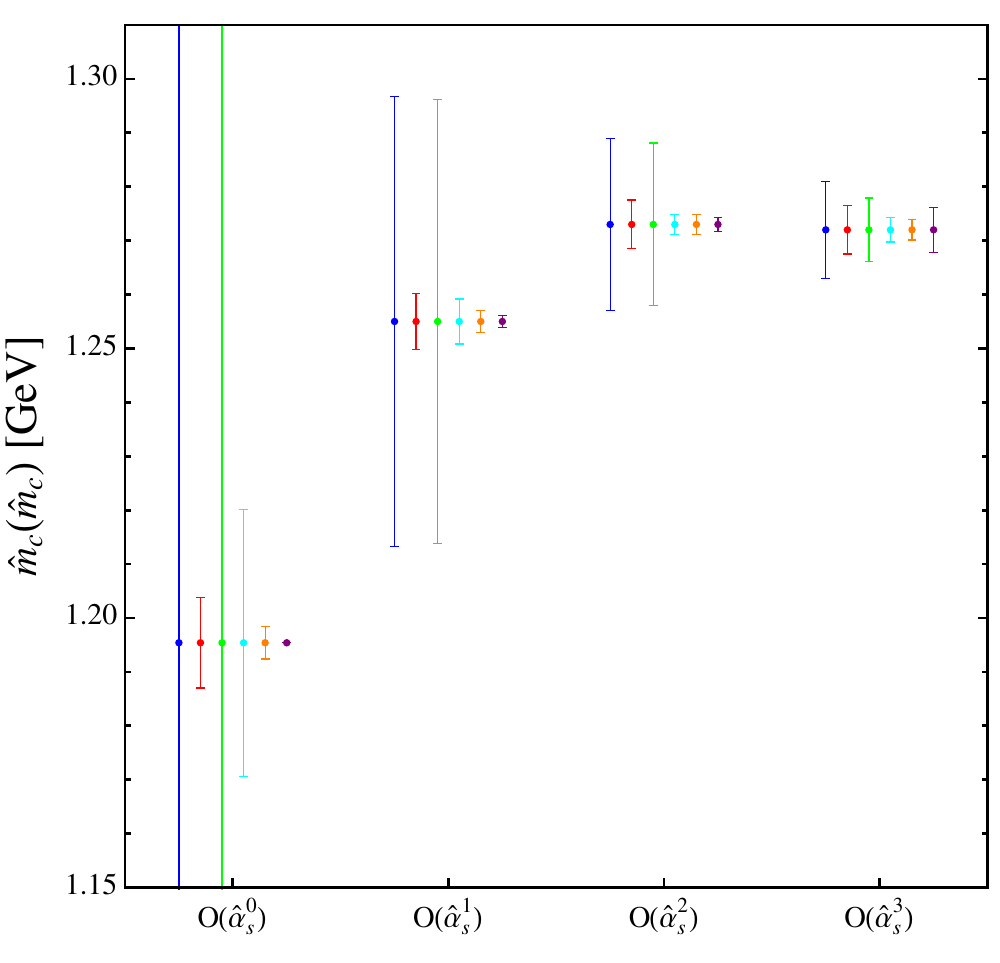}
\caption{Error budget for $\hat{m}_c(\hat{m}_c)$ when
determined from the pair of moments $({\cal M}_0, {\cal M}_2)$ at different orders in $\hat\alpha_s$.}
\label{Fig:cmass-an}
\end{figure}

Figure~\ref{Fig:cmass-an} shows the total uncertainty and the error breakdown as a function of the order in the perturbative expansion.
The experimental input error from the electronic partial widths is shown in red.
For the truncation error (in green) we used a somewhat more conservative estimate than simply taking the last available term.
As for the gluon condensate, we take the entire estimated contribution as the uncertainty (in orange).
We also show the parametric uncertainty (in purple) from $\hat\alpha_s(M_Z) = 0.1182 \pm 0.0016$.
Finally, we want to allow for the possibilities of larger-than-expected duality violations 
(while the assumption of quark-hadron duality in a finite region is much weaker than local duality, it still lacks complete rigour)
or high-order terms in the operator product expansion (OPE), given that the charm quark mass is dangerously close to the hadronic scale.
Therefore, we use $e^+ e^- \to$~hadrons electro-production data as our control of the method,
and compare the constraints on the continuum region from internal consistency by simultaneously considering different sum rules 
on the one hand with actual measurements on the other.
We add the difference between the two constraints
and the uncertainty in the experimental continuum data itself in quadrature (in cyan).

\begin{figure}
\centering
\includegraphics[width=.6\textwidth]{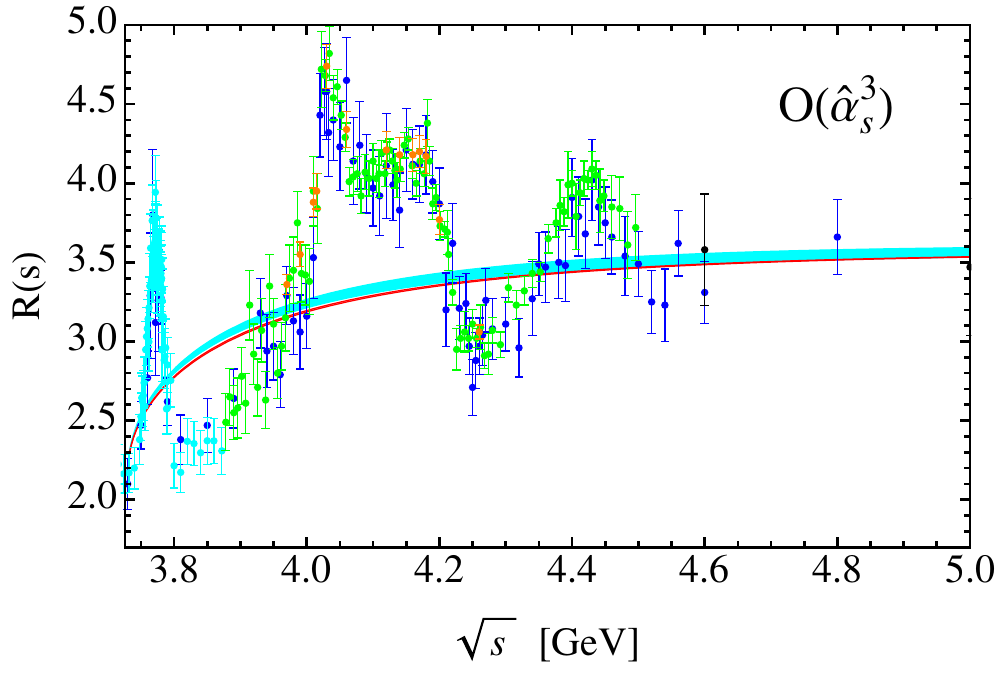}
\caption{Data for the ratio $R(s)$ for $e^+e^- \rightarrow $~hadrons relative to muons in the charm threshold region 
from Crystal Ball~\cite{Osterheld:1986hw}, BES~\cite{Bai:2001ct,Ablikim:2006mb}, and CLEO~\cite{CroninHennessy:2008yi}. 
For the full (orange) curve we have used $\lambda_3$ determined from the pair of moments ${\cal M}_0$ and ${\cal M}_2$,
while for the cyan band we have used $\lambda_3$ determined from data.}
\label{Rcan}
\end{figure}

Let us now have a closer look at our treatment of the continuum region.
Relativistic sum rules are those for rather low values in the moment number.
They can not reproduce the resonances, wiggles and other details in the continuum.
Neither would that be necessary because we only need integrals representing weighted averages.
Therefore, we make the simple continuum {\em ansatz\/}~\cite{Erler:2002bu},
\begin{equation}
R_c^{\rm cont}(s) = \frac{4}{3} \lambda_1(s) \sqrt{1 - \frac{4\, \hat m_q^2 (2 M_D)}{s'}} 
\left[ 1 + \lambda_3 \frac{2\, \hat m_c^2(2 M_D)}{s'} \right],
\label{ansatz}
\end{equation}
with
\begin{equation}
s' \equiv s + 4 [ \hat{m}_c^2(2M_D) - M_D^2 ],
\end{equation}
where $\lambda_1(s)$ is the known asymptotic behaviour and $\lambda_3$ is the only free parameter.
We expect $\lambda_3 \approx 1$, the value corresponding to the threshold behaviour of a fermion.
$R_c^{\rm cont}(s)$ smoothly interpolates between the onset of open charm production and perturbative QCD.
Considering ${\cal M}_0$ and ${\cal M}_2$ to constrain $m_c$ and $\lambda_3$ simultaneously,
we find $\lambda_3 = 1.23 \pm 0.06$, confirming our expectation and corresponding to the orange curve in Figure~\ref{Rcan}.
On the other hand, removing the background from light quarks and the small singlet contribution from the data 
provided by Crystal Ball~\cite{Osterheld:1986hw}, BES~\cite{Bai:2001ct,Ablikim:2006mb}, and CLEO~\cite{CroninHennessy:2008yi}, 
we find that the value $\lambda_3 = 1.34 \pm 0.17$ would reproduce the integral over data in the energy interval up to about 5~GeV.
Alternatively, we can subtract the background by performing a fit to the normalisation of the sub-continuum data
to perturbative QCD which amounts in $\lambda_3 = 1.15 \pm 0.16$.
It is remarkable and reassuring that all three values agree well within errors.

\begin{figure}
\centering
\includegraphics[width=.6\textwidth]{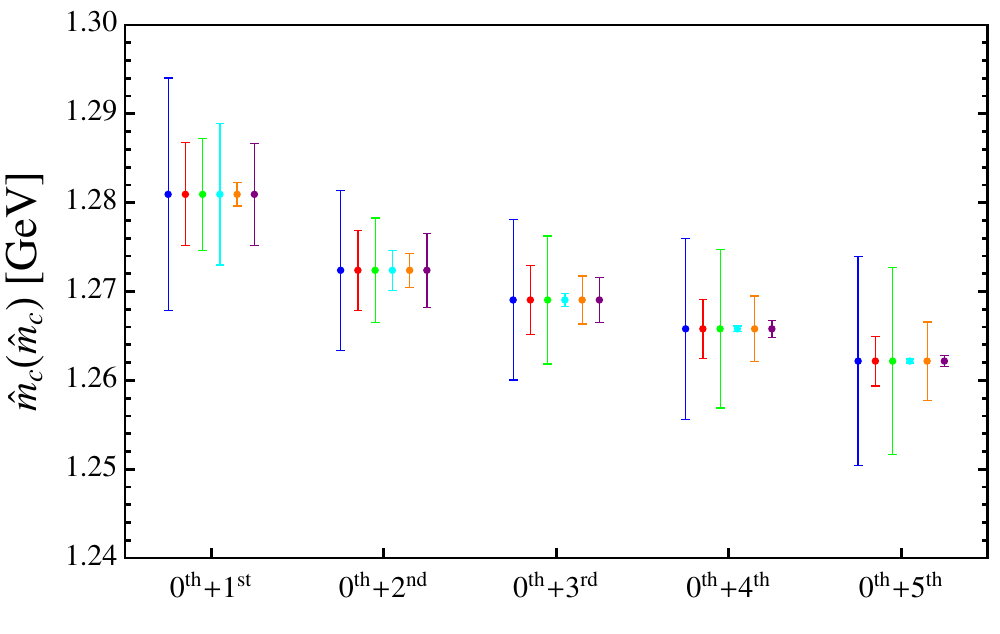}
\caption{$\hat{m}_c(\hat{m}_c)$ using different combinations of moments and the corresponding uncertainties. 
The color coding is as in Fig.~\ref{Fig:cmass-an}.
Notice, that all determinations are mutually consistent within our error estimates.}
\label{Fig:cmass}
\end{figure}

Other fits are possible.
If one considers ${\cal M}_0$ and ${\cal M}_1$ one faces a larger error from the continuum region 
as can be seen from the cyan coloured error bar in Figure~\ref{Fig:cmass}.
Or one can use the pairs $({\cal M}_0, {\cal M}_3)$ or $({\cal M}_1, {\cal M}_2)$, 
but in these cases the OPE truncation is of concern and indeed the gluon condensate is already contributing a larger error.
Another possibility is to consider three or four moments simultaneously with some assumption regarding
their mutual theoretical correlations.
{\em E.g.\/}, one can assume a common non-vanishing correlation coefficient between several moments,
and adjust it to enforce a value of the overall $\chi^2$ that equals the effective number of degrees of freedom. 
All of these and other options differ by no more than 4~MeV in the extracted $\hat m_c(\hat m_c)$.

\begin{figure}[b]
\centering
\includegraphics[width=0.79\textwidth]{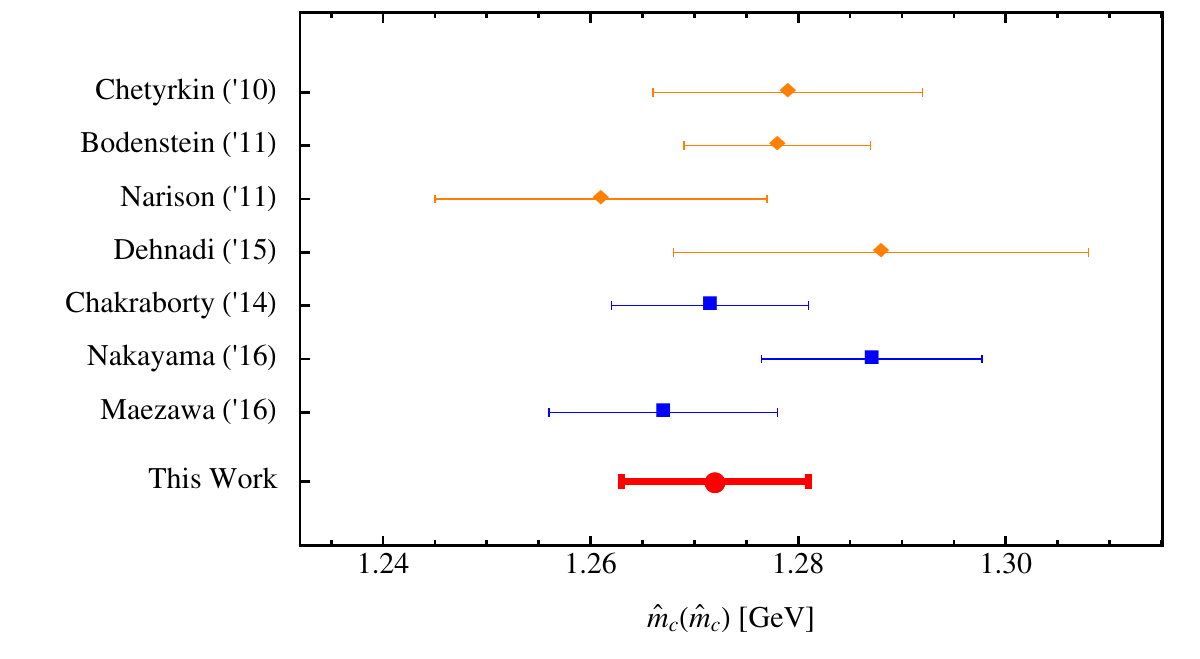}
\caption{Recent charm quark mass determinations (see text for details and references).}
\label{Fig:Comparison}
\end{figure}

\section{Conclusions and outlook}

We find very good agreement within the stated errors with other sum rule determinations of $\hat m_c(\hat m_c)$ using
${\cal M}_1$~\cite{Chetyrkin:2010ic,Dehnadi:2015fra}, ${\cal M}_2$~\cite{Bodenstein:2011ma}, and higher moments~\cite{Narison:2011xe}.
Our result also agrees very well with those from lattice simulations based on the pseudo-scalar current~\cite{Chakraborty:2014aca},
domain-wall fermions~\cite{Nakayama:2016atf}, and staggered quarks~\cite{Maezawa:2016vgv}.

In summary, we used a physically motivated continuum {\em ansatz\/} that reproduces the experimental data,
both the normalisation and --- non-trivially and very importantly --- the moment dependence, with a single adjustable parameter.
A non-parametric uncertainty of less than 0.7\% in a quantity that lives close to 1~GeV and is calculated perturbatively may seem optimistic.
But one should rather view this as a 3\% uncertainty in the difference of one half of the $J/\psi$ mass and $\hat m_c(\hat m_c)$.
Extrapolating this to the case of the bottom quark, which is work currently in preparation,
we would expect about 15~MeV uncertainty in $\hat m_b(\hat m_b)$ from the pair of moments $({\cal M}_0, {\cal M}_2)$.

\section*{Acknowledgments}

This work has been supported by DFG through the Collaborative Research Center 
``The Low-Energy Frontier of the Standard Model'' (SFB~1044). 
JE is supported by CONACyT (M\'exico) project 252167--F and acknowledges support from the Mainz cluster of excellence PRISMA.

\end{document}